\documentclass[10pt]{iopart}
\usepackage{iopams}
\usepackage{graphicx}
\usepackage{hyperref}
\newcommand{\mmm}{\mathbf m}
\newcommand{\rrr}{\mathbf r}
\newcommand{\hhh}{\mathbf h}
\newcommand{\AAA}{\mathbf A}

\begin{document}
\title{Unified theory of spin and charge in a ferromagnet}
\author{Oleg Tchernyshyov}
\address{William H Miller III Department of Physics and Astronomy and Institute for Quantum Matter, Johns Hopkins University, Baltimore, MD 21218, USA}

\begin{abstract}
We derive a unified theory of spin and charge degrees of freedom in a ferromagnet. The spin-transfer torque and spin electromotive force are examined from the coarse-grained perspective of collective coordinates. The resulting equations of motion reflect a balance of conservative, gyroscopic (Berry-phase), and dissipative forces. We then expand the space of collective coordinates by adding the electric charge. The adiabatic spin-transfer torque and spin emf turn out to be a gyroscopic force; their nonadiabatic counterparts are a dissipative force. 
\end{abstract}

\submitto{\JPCM}
\maketitle

\tableofcontents

\section{Introduction}
\label{sec:intro}

A spin-polarized electric current flowing through a ferromagnet induces spin dynamics in it, an effect known as the spin-transfer torque. The basic idea was formulated in 1996 for by Slonczewski \cite{Slonczewski:1996} and Berger \cite{Berger:1996} for two discrete magnets with spin directions given by unit vectors $\mmm_1$ and $\mmm_2$. The spins of conduction electrons passing through the first magnet get polarized along $\mmm_1$ and, upon entering the second magnet, switch their polarization to $\mmm_2$. The change in the spin orientation indicates a transfer of angular momentum between the magnets and conduction electrons. The reaction torque applied by the current to the magnets is the spin-transfer torque. It can be used to reverse the direction of magnetization in one of the ferromagnets \cite{Ralph:2008}. 

The notion of the spin-transfer torque was extended to ferromagnets with continuously varying magnetization $\mmm(\rrr)$ by Bazaliy \emph{et al.} \cite{Bazaliy:1998} (and anticipated 20 years prior by Berger \cite{Berger:1978}). In their theory, conduction electrons flowing through the ferromagnet instantenously align their spins with the local magnetization $\mmm(x)$, hence the name \emph{adiabatic} spin-transfer torque. Tatara and Kohno \cite{Tatara:2004} and Zhang and Li \cite{Zhang:2004} argued that a slight misalignment generates what has become known as the \emph{nonadiabatic} spin-transfer torque.  Tserkovnyak \emph{et al.} \cite{Tserkovnyak:2006, Tserkovnyak:2009} suggested an alternative, dissipative origin of the nonadiabatic spin-transfer torque. Garate \emph{et al.} \cite{Garate:2009} provided a microscopic derivation of this term as a dissipative force. 

The spin-transfer torques enable a practical way of manipulating magnetic solitons with an electric current \cite{Kent:2015}. Their physics was reviewed in depth in a 2008 special issue of the Journal of Magnetism and Magnetic Materials \cite{Ralph:2008}, with input from experts in both theory \cite{Berkov:2008, Tserkovnyak:2008, Ohno:2008, Haney:2008} and experiment \cite{Katine:2008, Sun:2008, Silva:2008, Beach:2008}. 

What, then, compels us to write another theory paper on this mature topic? The following quote from the editorial introduction to the 2008 special issue \cite{Ralph:2008} reveals a lack of consensus among experts about the very basics of the spin-transfer torques: 
\begin{quote}
    ``Much of the debate on the theoretical description of current-induced domain wall motion is associated with how to describe the damping and whether there is an additional torque in the direction of the non-adiabatic torque... that arises from the same processes that lead to damping... This contribution is extensively discussed by Tserkovnyak, Brataas, and Bauer (this issue). One technical aspect, whether the adiabatic spin transfer torque can be derived from an energy functional, is addressed in the article by Haney, Duine, N{\'u}{\~n}ez, and MacDonald (this issue). One of the present authors believes that their argument is incorrect, but disagreement is one of the things that makes an issue an open question.''
\end{quote}
As far as we know, the controversies surrounding the nature of the spin-transfer torques have not been resolved to everyone's satisfaction \cite{experts:private-comm}. We hope to bring some clarity to these issues. 

To get a clearer picture of this problem, it helps to expand its scope. The adiabatic spin-transfer torque has an inverse effect fully elucidated in 2007 by Barnes and Maekawa \cite{Barnes:2007} (and anticipated 20 years prior by Berger \cite{Berger:1986} and Volovik \cite{Volovik:1987}). Dynamical magnetization exerts an electromotive force (emf) on conduction electrons. Although the two effects are intimately related, they are usually treated separately, with the magnetic and electric degrees of freedom described at different levels of detail; a unified description treating spin and charge on the same footing is insightful but rare \cite{Tserkovnyak:2009, Brataas:2017}. Both effects can be derived from a common set of rules expressed, e.g., by a unified Lagrangian. 

The role of electric charge in spin-transfer torques has of course been studied by many authors (see Ralph and Stiles \cite{Ralph:2008} and references therein). However, most of the approaches are based on specific microscopic models, involve many complex degrees of freedom, and rely on specialized techniques. We wish to avoid that and will keep the number of physical variables to a bare minimum. An effective description of the low-energy dynamics of a magnetic soliton is achieved through a coarse-graining procedure based on collective coordinates \cite{Tretiakov:2008} such as the position and orientation of a domain wall in a magnetic wire. In the unified approach, we treat the electric charge as one of the collective coordinates of the combined physical system. Doing so allows us to describe the dynamics for the combined system in a compact and uniform way. The resulting equations of motion (\ref{eq:eom-unified}) reflect the balance of generalized forces acting on the soliton and on the electric charge. 

Once the dynamics of the combined system is expressed in this compact way, we examine the resulting equations of motion and analyze the nature of the relevant terms. Generalized forces in these equations come in three distinct types: conservative, gyroscopic, and dissipative. The adiabatic spin-transfer torque and its inverse effect, the spin emf, together constitute a gyroscopic force similar in nature to the Lorentz and Coriolis forces. The nonadiabatic spin-transfer torque and its electric counterpart are a dissipative force. These identifications can only be made by considering both spin and charge degrees of freedom on the same footing.

The rest of the paper is organized as follows. We begin by classifying forces relevant to ferromagnetic solitons (conservative, dissipative, and gyroscopic) in Sec.~\ref{sec:forces}. In Sec.~\ref{sec:ASTT} we show that the adiabatic spin-transfer torque is not a conservative force and argue that, with electric charge included as a physical degree of freedom, this torque should be viewed as a gyroscopic force. A unified treatment of spin and charge degrees of freedom of the combined system (ferromagnet and electric circuit) is worked out in Sec.~\ref{sec:unification}, yielding the geometric action for both the spin-transfer torque and its inverse effect, the spin emf. In Sec.~\ref{sec:NASTT} the nonadiabatic torque is shown to be neither conservative, nor gyroscopic, but dissipative. Appendices contain some helpful derivations: equations of motion for an electric circuit are derived from the Aharonov-Bohm phase in \ref{app:electric-circuit-geometric-phase}; the spin-tramsfer torque is derived from the spin Berry phase in \ref{app:ferromagnet-geometric-phase}. 

\section{A classification of forces}
\label{sec:forces}

The arguments that follow rely on a classification of forces into three types: conservative, dissipative, and gyroscopic. The reader is undoubtedly familiar with the first of these, but much less so with the last. Let us define these types. 

Magnetic dipoles in ferromagnets come from spins that act like microscopic gyroscopes. Anyone who played with a gyroscope or a fast-spinning top knows that their dynamics are a bit weird: pushed in one direction, the axis of a gyroscope precesses in the orthogonal direction. This weirdness translates to the dynamics of solitons in a ferromagnet. When a domain wall is pushed, it turns; and when it is twisted, it moves. Spins lack inertia, so their precession begins instantaneously, as soon as an external torque is applied. This translates to the absence of kinetic energy in a ferromagnetic soliton. At the most basic level, its equations of motion lack the acceleration term familiar from Newton's second law \footnote{Inertia in a ferromagnetic soliton can occur as an emergent phenomenon. The D{\"o}ring mass can arise after a coarse-graining procedure that eliminates some high-energy degrees of freedom and endows the remaining low-energy degrees of freedom with kinetic energy \cite{Doring:1948, Rado:1950, Rado:1951, Saitoh:2004}.} 

A ferromagnetic soliton can be parametrized in terms of a few collective coordinates 
\begin{equation}
q \equiv \{q^1, q^2, \ldots, q^N\}   
\label{eq:q}
\end{equation}
(see Sec.~\ref{sec:ASTT-is-not-conservative} below for an example). Their dynamics can be obtained from the action 
\begin{equation}
S = 
- \int U \, dt
+ \int A_i \, dq^i.
\label{eq:S-soliton}
\end{equation}
The first term in Eq.~(\ref{eq:S-soliton}) is a dynamical action coming from potential energy $U(q)$. The second term is a geometric action \cite{Shapere:1989}, also known as the Aharonov-Bohm phase and Berry phase in specific contexts; it includes a gauge field, or connection, $A_i(q)$. Minimizing the action yields an equation of motion for each coordinate $q^i$, 
\begin{equation}
0 = \frac{\delta S}{\delta q^i(t)} 
= 
- \frac{\partial U}{\partial q^i}
+ G_{ij} \dot{q}^j.
\label{eq:eom-collective-coordinates}
\end{equation}
Summation is implied over doubly repeated indices. Here 
\begin{equation}
G_{ij} = 
\frac{\partial A_j}{\partial q^i}
- 
\frac{\partial A_i}{\partial q^j}.
 \label{eq:G-curl-A}
\end{equation}
is an antisymmetric tensor, $G_{ij} = - G_{ji}$. 

The equation of motion (\ref{eq:eom-collective-coordinates}) expresses the balance of generalized forces conjugate to coordinate $q^i$. The first term is a \emph{conservative} force derived from the potential energy $U(q)$. The second term is a \emph{gyroscopic} force, a name that originated with Kelvin \cite{Thomson:1879, Krechetnikov:2007}. Berry referred to such forces as \emph{geometric} \cite{Berry:1993}. 

Familiar examples of a gyroscopic force are the Lorentz force $\mathbf F = e \, \dot{\mathbf r} \times \mathbf B$ acting on an electric charge moving in a magnetic field $\mathbf B$ and the Coriolis force $\boldsymbol \Omega$, $\mathbf F = 2 m \dot{\mathbf r} \times \boldsymbol \Omega$ in a reference frame rotating at an angular frequency. The gyroscopic tensor in these cases is $G_{ij} = e \epsilon_{ijk} B_k$ and $G_{ij} = 2 m \epsilon_{ijk} \Omega_k$, respectively.  

The gyroscopic tensor satisfies an important constraint known as the Bianchi identity \cite{Clarke:2008, Tchernyshyov:2015}: 
\begin{equation}
\rho_{ijk} 
\equiv \frac{\partial G_{jk}}{\partial q^i}
+ \frac{\partial G_{ki}}{\partial q^j}
+ \frac{\partial G_{ij}}{\partial q^k}
= 0.
\label{eq:Bianchi}
\end{equation}
It follows directly from Eq.~(\ref{eq:G-curl-A}). In the case of the Lorentz force, the Bianchi identity (\ref{eq:Bianchi}) expresses the absence of magnetic charges:
\begin{equation}
4\pi \rho \equiv \nabla \cdot \mathbf B = 0.    
\end{equation}
It is therefore helpful to think of the gyroscopic tensor $G_{ij}$ as of a magnetic field in the manifold of collective coordinates \cite{Shapere:1989}. The Bianchi identity (\ref{eq:Bianchi}) then expresses the absence of sources or sinks for the magnetic field.

It is well-known that the work of a conservative force can be expressed as an increment of its potential energy, 
\begin{equation}
-\frac{\partial U}{\partial q^i} dq^i = -dU(q).    
\end{equation}
The work of a gyroscopic force is strictly zero, 
\begin{equation}
G_{ij} \dot{q}^j dq^i 
= G_{ij} \dot{q}^j \dot{q}^i dt 
= 0.
\end{equation}
The last step follows from the antisymmetry of the gyroscopic tensor. 

One more type of forces relevant to the dynamics of magnetic solitons is \emph{dissipative}, or viscous. Like the gyroscopic force, it is proportional to generalized velocities, $F_i = - \Gamma_{ij}\dot{q}^j$. However, the proportionality coefficients are symmetric, $\Gamma_{ij} = \Gamma_{ji}$, and form a positive-definite tensor. That ensures that the work of a dissipative force is always negative or zero: 
\begin{equation}
- \Gamma_{ij}\dot{q}^j dq^i 
= - \Gamma_{ij}\dot{q}^j \dot{q}^i dt 
\leq 0.
\end{equation}
A dissipative force cannot be obtained from minimization of some action. It is an emergent force, arising from interaction of a macroscopic object with numerous microscopic degrees of freedom. In thermodynamics, it plays a major role in the relaxation of the physical system toward thermal equilibrium. In that context, the dissipative force can be obtained from the Rayleigh dissipation function \cite{LL:5},
\begin{equation}
F_i = -\frac{\partial R}{\partial \dot{q}^i},
\quad
R = \frac{1}{2} \Gamma_{ij} \dot{q}^i \dot{q}^j.
\label{eq:Rayleigh}
\end{equation}

\section{Adiabatic spin-transfer torque}
\label{sec:ASTT}

\subsection{Conservative? No}
\label{sec:ASTT-is-not-conservative}

We first show that the (adiabatic) spin-transfer torque is not a conservative force; therefore, it cannot be derived from an energy functional. 

Consider a ferromagnetic wire whose spin distribution is described by a vector field $\mmm(t,x)$ of unit length. Its dynamics is described by the Landau--Lifshitz equation, equating the rate of change of the angular momentum with torque \cite{Bazaliy:1998, Tserkovnyak:2008}: 
\begin{equation}
\mathcal S \, \partial_t \mmm = 
- \mmm \times \frac{\delta U[\mmm]}{\delta \mmm}
- \frac{\hbar I}{2e} \partial_x \mmm.
\label{eq:LL-1d-torques}
\end{equation}
Here $\mathcal S$ is the (scalar) spin density per unit length; $U[\mmm(x)]$ is an energy functional; the functional derivative $- \delta U[\mmm]/\delta \mmm(x) \equiv \hhh(x)$ is the effective magnetic field; $I$ is the electric current flowing through the wire; $e$ is the elementary electric charge. We postpone the discussion of dissipative terms (such as the Gilbert damping) until Sec.~\ref{sec:NASTT-is-dissipative}. The last term on right hand-side of Eq.~(\ref{eq:LL-1d-torques}) is the density of the spin-transfer torque. For simplicity, we assume that the electric current in the ferromagnet is fully spin-polarized. 

It is convenient to restate the Landau--Lifshitz equation in a modified form, which features generalized forces instead of torques. To that end, we take a cross product of Eq.~(\ref{eq:LL-1d-torques}) with $\mmm$ and use some vector algebra to obtain the following equation: 
\begin{equation}
- \frac{\delta U[\mmm]}{\delta \mmm} 
- \mathcal S \, \partial_t \mmm \times \mmm
- \frac{\hbar I}{2e} \partial_x \mmm \times \mmm = 0.
\label{eq:LL-1d-forces}
\end{equation}
The modified Landau--Lifshitz equation (\ref{eq:LL-1d-forces}) expresses equilibrium of generalized forces acting on the magnetization field. The first term is evidently a conservative force derived from an energy functional. The second term, responsible for precessional dynamics of spins in a ferromagnet, is a gyroscopic force (see Sec.~\ref{sec:ASTT-is-gyroscopic} for a precise definition). Demystifying the last force in Eq.~(\ref{eq:LL-1d-forces}), related to the spin-transfer torque, is one of our goals.  

If the spin-transfer torque were a conservative force then the last term in Eq.~(\ref{eq:LL-1d-forces}) could be expressed as the functional derivative of a spin-transfer energy $U^\mathrm{ASTT}[\mmm(x)]$. We shall see that this assumption leads to a contradiction \cite{Dasgupta:2018}. 

For an infinitesimal variation $\delta \mmm(x)$ of the magnetization field, the first variation of this energy would be
\begin{equation}
\delta U^\mathrm{ASTT} = \int dx \, 
\frac{\hbar I}{2e} (\partial_x \mmm \times \mmm)
\cdot \delta \mmm. 
\label{eq:delta-U-ST-m}
\end{equation}
Let us consider a specific example of a domain wall in a magnetic wire with easy-axis anisotropy. It has two uniform ground states $\mmm(x) = (\pm 1, 0, 0)$ and domain-wall solutions interpolating between them,
\begin{equation}
\mmm(x,X,\Psi) = 
(\pm \tanh{\frac{x-X}{\lambda}}, \,
\mathrm{sech}{\frac{x-X}{\lambda}} \cos{\Psi}, \,
\mathrm{sech}{\frac{x-X}{\lambda}} \sin{\Psi}).   
\label{eq:domain-wall-solution}
\end{equation}
The width of the domain wall $\lambda$ is determined by the competition of exchange and anisotropy energies. The free parameters $X$ and $\Psi$ are collective coordinates of a domain wall defining its position in the wire and its azimuthal orientation, respectively. The energy of a domain wall is independent of them, reflecting the global symmetries of translation and axial spin rotation. For a domain wall solution (\ref{eq:domain-wall-solution}), the spin-transfer energy should be a function of $X$ and $\Psi$. Expressing the first variation of the spin-transfer energy (\ref{eq:delta-U-ST-m}) in terms of $\delta X$ and $\delta \Psi$ yields $\delta U^\mathrm{ASTT} = \pm (\hbar I/e) \delta \Psi$, from which we obtain 
\begin{equation}
U^\mathrm{ASTT}(X,\Psi) = \pm \frac{\hbar I}{e} \Psi.
\label{eq:U-ST-X-Psi}
\end{equation}
This result is quite intuitive: the spin-polarized current applies no force to a domain wall, $-\partial U^\mathrm{ASTT}/\partial X = 0$, but exerts a torque $-\partial U^\mathrm{ASTT}/\partial \Psi = \mp \hbar I/e$. 

The problem with the spin-transfer energy (\ref{eq:U-ST-X-Psi}) is that it is not single-valued. If we rotate the plane of the domain wall once, incrementing $\Psi$ by $2\pi$, then the spin-transfer energy is incremented by $\pm 2\pi \hbar I/e$, indicating that the work done by the spin-transfer torque around a closed loop in the configuration space $(X,\Psi)$ is nonzero. Thus the (adiabatic) the spin-transfer torque is not a conservative force. 

\subsection{Gyroscopic? Yes}
\label{sec:ASTT-is-gyroscopic}

What kind of a force is it then? The language of collective coordinates provides a simple way to answer that question. Eq.~(\ref{eq:LL-1d-forces}) translates into equations of motion for collective coordinates,
\begin{equation}
- \frac{\partial U}{\partial q^i}
+ G_{ij}\dot{q}^j
+ F_i^\mathrm{ASTT} = 0.
\label{eq:eom-collective-coordinates-with-ASTT}
\end{equation}
The adiabatic spin-transfer torque induces a new generalized force 
\begin{equation}
F_i^\mathrm{ASTT} = - \frac{\hbar I}{2e} 
\int dx \, 
\mmm \cdot 
\left(
\frac{\partial \mmm}{\partial q^i} 
\times 
\partial_x \mmm
\right).     
\end{equation}
The nature of this force comes into clear focus if we express the electric current as the velocity of another physical variable, $I = \dot{Q}$. Here $Q(t)$ is the amount of electric charge that has flown through (any part of) the ferromagnetic wire since $t=0$. We thus obtain $F_i^\mathrm{ASTT} = G_{iQ} \dot{Q}$, a force linear in the generalized velocity $\dot{Q}$. The equation of motion for collective coordinates of a soliton (\ref{eq:eom-collective-coordinates}) now reads 
\begin{equation}
- \frac{\partial U}{\partial q^i}
+ G_{ij}\dot{q}^j
+ G_{iQ} \dot{Q} = 0.
\label{eq:eom-collective-coordinates-with-Q}
\end{equation}

The remarkable similarity between the second and third terms of Eq.~(\ref{eq:eom-collective-coordinates-with-Q}) suggests that the spin-transverse force should be viewed as a gyroscopic force if we treat the electric charge $Q$ as one of the collective coordinates. The new gyroscopic coefficients are 
\begin{equation}
G_{iQ} = 
- \frac{\hbar}{2e} 
\int dx \, 
\mmm \cdot 
\left(
\frac{\partial \mmm}{\partial q^i} 
\times 
\partial_x \mmm 
\right).
\label{eq:G-i-Q}
\end{equation}
For domain wall in a ferromagnetic wire (\ref{eq:domain-wall-solution}), $G_{XQ} = 0$ and $G_{\Psi Q} = \mp \hbar/e$. 

\section{Unification of spin and charge}
\label{sec:unification}

\subsection{New degrees of freedom: electric charge and magnetic flux}
\label{sec:new-degrees-of-freedom-Q-Phi}

The uniformity, with which collective coordinates $q^j$ and electric charge $Q$ enter the equations of motion (\ref{eq:eom-collective-coordinates-with-Q}), suggests that these variables ought to be treated on an equal footing. This compels us to promote electric charge $Q$ from an auxiliary variable expressing the strength of a background electric current $I = \dot{Q}$ to a physical degree of freedom. 

This logical leap has immediate implications. The charge degree of freedom has its own gyroscopic forces proportional to the velocities of other variables. The dynamics of magnetization, expressed by time-dependent collective coordinates of a magnetic soliton $q(t)$, produces a generalized force in the charge channel, i.e., an emf $F_Q = G_{Qi} \dot{q}^i$, where the gyroscopic coefficients $G_{Qi}$ are already fixed by the antisymmetry of the gyroscopic tensor, $G_{Qi} = - G_{iQ}$. For example, for a domain wall in one dimension considered in a previous section, $G_{QX} = -G_{XQ} = 0$ and $G_{Q\Psi} = -G_{\Psi Q} = \pm \hbar/e$. Thus a precessing domain wall creates a spin emf $\mathcal E^\mathrm{ASTT} = \pm \hbar \dot{\Psi}/e$ in accordance with Barnes and Maekawa \cite{Barnes:2007}. 

To complete the dynamical description of our extended system, we have to work with a one-dimensional ferromagnet forming a closed loop, i.e., an electric circuit. Endowing the circuit with a battery supplying an emf $V$ and a magnetic flux $\Phi$ enclosed by the loop yields the following equation of motion for charge $Q$: 
\begin{equation}
- \dot{\Phi} + V + G_{Qi} \dot{q}^i = 0.
\label{eq:eom-Q}
\end{equation}
The first term is the emf of Maxwell's induction, which is evidently another instance of the gyroscopic force with the gyroscopic coefficient $G_{Q\Phi} = - 1$. By the familiar logic, we add the magnetic flux $\Phi$ as one more physical variable. Assuming that the magnetic flux $\Phi$ is induced solely by the electric current flowing in the circuit, we have the relation $\Phi = \ell \dot{Q}$, where $\ell$ is the self-induction of the circuit. We rewrite it as
\begin{equation}
\dot{Q} - \Phi/\ell = 0,  
\label{eq:eom-Phi}
\end{equation}
so that the first term appears as a gyroscopic force $G_{\Phi Q} \dot{Q}$ with $G_{\Phi Q} = - G_{Q \Phi} = 1$. Then the second term is a conservative force $- \partial U^\mathrm{MS}/\partial \Phi$ from the magnetostatic energy $U^\mathrm{MS} = \Phi^2/2\ell$. 

\subsection{Unification of spin and charge}

Treating the magnetic and electric degrees of freedom on an equal footing allows us to express their dynamics in a simple and elegant way. Labeling the electric charge and magnetic flux as two more collective coordinates,
\begin{equation}
q = \{q^1, \ldots, \, q^N, \, q^{N+1} = Q, \, q^{N+2} = \Phi\},  
\label{eq:collective-coordinates-extended}
\end{equation}
allows us to write the three equations (\ref{eq:eom-collective-coordinates-with-Q}), (\ref{eq:eom-Q}), and (\ref{eq:eom-Phi}) in one line: 
\begin{equation}
- \frac{\partial W}{\partial q^\alpha}
+ G_{\alpha\beta} \dot{q}^\beta 
= 0.
\label{eq:eom-unified}
\end{equation}
We use Greek indices $\alpha = 1, \ldots, N+2$ to label the extended set of coordinates; again, doubly repeated indices are summed over. The net potential energy $W$ includes contributions from both the magnet and the circuit: 
\begin{equation}
W(q) = U(q^1, \ldots, q^N) + \frac{\Phi^2}{2\ell} - VQ.
\end{equation}

\subsection{Geometric action for the adiabatic spin-transfer torque}
\label{sec:action-lagrangian}

The action giving rise to the equations of motion (\ref{eq:eom-unified}) is 
\begin{equation}
S = \int A_\alpha \, dq^\alpha - \int W \, dt 
= \int (A_\alpha \dot{q}^\alpha - W) dt .
\label{eq:S-extended}
\end{equation}
The term $\int A_\alpha(q) dq^\alpha$ is the geometric action (Berry phase) containing the gauge potentials $A_\alpha(q)$, whose curl gives the extended gyroscopic tensor:
\begin{equation}
\frac{\partial A_\beta}{\partial q^\alpha}
- \frac{\partial A_\alpha}{\partial q^\beta}
= G_{\alpha\beta}.
\label{eq:G-curl-A-extended}
\end{equation}
It is helpful to think of $G_{\alpha\beta}$ as of a magnetic field in configuration space (\ref{eq:collective-coordinates-extended}). Then the gyroscopic force $G_{\alpha\beta} \dot{q}^\beta$ is the Lorentz force acting on a particle moving in that space; the geometric part of the action (\ref{eq:S-extended}) is the Aharonov-Bohm phase in this analogy. 

The geometric part of the action has been extensively discussed previously for the ferromagnet alone \cite{Haney:2008, Clarke:2008, Tchernyshyov:2015}, so we will not dwell on that. Our immediate goal is to obtain that part of the geometric action which is responsible for the spin-transfer torque and the spin emf. The relevant components of the gyroscopic tensor $G_{iQ}$ (\ref{eq:G-i-Q}) are independent of the electric charge. We may therefore use the following gauge potential: 
\begin{equation}
A_i = - G_{iQ} Q,
\quad
A_Q = A_\Phi = 0.
\end{equation}
The corresponding geometric action is 
\begin{equation}
S_g^\mathrm{ASTT} 
= \int A_\alpha \dot{q}^\alpha dt
= \int dt \int dx \, \frac{\hbar Q}{2e} \dot{q}^i \,
\mmm \cdot 
\left(
\frac{\partial \mmm}{\partial q^i}
\times \partial_x \mmm  
\right).
\end{equation}
With the aid of the chain rule,  $\partial_t\mmm(x,q(t)) = \dot{q}^i 
\partial \mmm/\partial q^i$, we obtain one of our main results, the geometric action encoding both the adiabatic spin-transfer torque and spin emf: 
\begin{equation}
S_g^\mathrm{ASTT} 
= \int dt \int dx \, \frac{\hbar Q}{2e} \,
\mmm \cdot 
\left(
\partial_t \mmm \times 
\partial_x \mmm
\right).
\label{eq:S-stt-smf-1+1}
\end{equation}

The adiabatic spin-transfer action (\ref{eq:S-stt-smf-1+1}) resembles very closely the Wess-Zumino action for a single spin \cite{Dasgupta:2018, Fradkin:1988}, with two major distinctions. First, the action for a single spin is written with an unphysical extra dimension, whereas the spatial dimension in Eq.~(\ref{eq:S-stt-smf-1+1}) is real. Second, the first variation of the Wess-Zumino action comes strictly from the boundary of the spacetime; in contrast, the presence of the time-dependent variable $Q(t)$ in the integrand of Eq.~(\ref{eq:S-stt-smf-1+1}) gives rise to a bulk term in $\delta S_g^\mathrm{ASTT}$. The functional derivatives of this action with respect to the magnetization field and the electric charge are, respectively, the adiabatic spin-transfer torque 
\begin{equation}
\mmm \times \frac{\delta S_g^\mathrm{ASTT}}{\delta \mmm} 
= - \frac{\hbar \dot{Q}}{2e} \partial_x \mmm,
\label{eq:STT-m}
\end{equation}
and the spin emf,
\begin{equation}
\frac{\delta S_g^\mathrm{ASTT}}{\delta Q}
= \int dx \, \frac{\hbar}{2e}
\mmm \cdot 
\left(
\partial_t \mmm \times 
\partial_x \mmm
\right)
= \int E_x \, dx.
\end{equation}
Here $E_x = \frac{\hbar}{2e} \mmm \cdot 
\left(
\partial_t \mmm \times 
\partial_x \mmm
\right)$ is Volovik's emergent electric field \cite{Volovik:1987}.

\section{Nonadiabatic spin-transfer torque}
\label{sec:NASTT}

Tatara and Kohno \cite{Tatara:2004} and Zhang and Li \cite{Zhang:2004} suggested that the spin-transfer torque has a nonadiabatic component, arising when the spins of conduction electrons moving through the ferromagnet are slightly misaligned with its local magnetization. The resulting Landau--Lifshitz equation then includes a ``$\beta$ term,''
\begin{equation}
\mathcal S \, \partial_t \mmm = 
- \mmm \times \frac{\delta U[\mmm]}{\delta \mmm}
- \frac{\hbar I}{2e} \partial_x \mmm
- \beta \frac{\hbar I}{e} \mmm \times \partial_x\mmm.
\label{eq:LL-1d-NASTT-torques}
\end{equation}
The nature of the $\beta$ term is also controversial \cite{Ralph:2008}. 

\subsection{Conservative? No}
\label{sec:NASTT-is-not-conservative}

First we establish that, with the electric current treated as a background, the $\beta$ term is not a conservative force. Working along the lines of Sec.~\ref{sec:ASTT-is-not-conservative}, we assume the existence of potential energy $U^\mathrm{NSTT}$ generating the nonadiabatic spin-transfer torque. Under an infinitesimal change of the magnetization field $\delta \mmm(x)$, the energy would be incremented by 
\begin{equation}
\delta U^\mathrm{NSTT} 
= \beta \frac{\hbar I}{e} 
\int dx \, \partial_x \mmm \cdot \delta \mmm
\end{equation}
For a uniform translation of a soliton by $\delta X$,  
\begin{equation}
\delta U^\mathrm{NSTT} 
= - \beta \frac{\hbar I}{e} \delta X
\int dx \, (\partial_x \mmm)^2. 
\end{equation}
Assume a ferromagnetic wire with the topology of a loop. Moving the soliton around the loop once, and thus returning the system to its original state, gives a nonzero increment of the hypothetical potential energy $U^\mathrm{NSTT}$, a contradiction.

\subsection{Gyroscopic? No}
\label{sec:NASTT-is-not-gyroscopic}

Next we make the electric current a dynamical degree of freedom with the electric charge $Q(t)$ as its coordinate. Being linear in the velocity $\dot{Q}$, the nonadiabatic spin-transfer torque could possibly be gyroscopic in nature. In that case, it would be derivable from a geometric term in the action (a Berry phase). We shall explore this possibility and conclude that it leads to an unphysical term in the Landau-Lifshitz equation. 

Assuming that the $\beta$ term is a gyroscopic force, its effect on magnetic collective coordinate $q^i$ would be quantified by the generalized force 
\begin{equation}
F_i^\mathrm{NSTT} = G^\mathrm{NSTT}_{iQ} \dot{Q},
\quad
G^\mathrm{NSTT}_{iQ}
= - \beta \frac{\hbar}{e}
\int dx \, 
\frac{\partial \mmm}{\partial q^i}
\cdot \partial_x \mmm.
\end{equation}
The inverse effect would be a spin-generated emf $\mathcal E = G^\mathrm{NSTT}_{Qi} \dot{q}^i$ with the gyroscopic coefficient $G^\mathrm{NSTT}_{Qi} = - G^\mathrm{NSTT}_{iQ}$. 

Before obtaining a gauge potential $A^\mathrm{NSTT}_\alpha$ responsible for the $\beta$ term, we have to make sure that the gyroscopic tensor $G^\mathrm{NSTT}_{\alpha\beta}$ satisfies the Bianchi identity (\ref{eq:Bianchi}). Let us look at the following triplet of coordinates: two magnetic ones $q^i$ and $q^j$ and the electric charge $Q$. We expect translational symmetry for the charge: no physical quantity should depend on the amount of charge that has flown through the magnetic wire. we would thus expect that the gyroscopic coefficient $G_{ij}$, which quantifies the gyroscopic force $G_{ij}\dot{q}^j$, is independent of $Q$. Under this assumption, the Bianchi identity is violated: 
\begin{equation}
\rho^\mathrm{NSTT}_{ijQ}
= \beta \frac{\hbar}{e}
\int dx \, 
\left(
\frac{\partial \mmm}{\partial q^i}
\cdot 
\partial_x 
\frac{\partial \mmm}{\partial q^j}
- \frac{\partial \mmm}{\partial q^j}
\cdot 
\partial_x 
\frac{\partial \mmm}{\partial q^i}
\right)
\neq 0.
\end{equation}

The Bianchi identity can be saved at the price of assuming a $Q$-dependent gyroscopic coefficient 
\begin{equation}
G^\mathrm{NSTT}_{ij} 
= -\beta \frac{\hbar Q}{e}
\int dx \,
\left(
\frac{\partial \mmm}{\partial q^i}
\cdot 
\partial_x 
\frac{\partial \mmm}{\partial q^j}
- \frac{\partial \mmm}{\partial q^j}
\cdot 
\partial_x 
\frac{\partial \mmm}{\partial q^i}
\right).
\end{equation}
The revised gyroscopic tensor can be obtained from the following gauge potential: 
\begin{equation}
A^\mathrm{NSTT}_Q = 0, 
\quad
A^\mathrm{NSTT}_i = 
\beta \frac{\hbar Q}{e}
\int dx \, 
\frac{\partial \mmm}{\partial q^i}
\cdot \partial_x \mmm.
\end{equation}

The geometric action for the nonadiabatic spin-transfer torque is 
\begin{equation}
S_g^\mathrm{NSTT} 
= \int A^\mathrm{NSTT}_\alpha \dot{q}^\alpha \, dt
= 
\int dt \int dx \, 
\beta\frac{\hbar Q}{e} \, \partial_t \mmm \cdot \partial_x \mmm.
\end{equation}
It is quite similar to the action (\ref{eq:S-stt-smf-1+1}) for the adiabatic spin-transfer torque, with one substantial difference: the topological density $\mmm \cdot (\partial_t \mmm \times \partial_x \mmm)$ is replaced with a nontopological quantity $\partial_t \mmm \cdot \partial_x \mmm$. As a result, the variation of $S_g^\mathrm{NSTT}$ will contain a bulk term proportional to the charge $Q$, in contrast to $S_g^\mathrm{ASTT}$ whose variation lacks such a term. The nonadiabatic spin-transfer torque derived from this action will have, aside from the familiar $\beta$ term, an unphysical charge-dependent component: 
\begin{equation}
\mmm \times \frac{\delta S_g^\mathrm{NSTT}}{\delta \mmm} 
= - \beta \frac{\hbar \dot{Q}}{e} \mmm \times \partial_x\mmm
- 2 \beta \frac{\hbar Q}{e} 
\mmm \times \partial_t \partial_x \mmm.
\end{equation}

We thus find that our attempt to describe the nonadiabatic spin-transfer torque as a gyroscopic force derivable from a geometric action has failed: the rigid structure of the gyroscopic force demands that the usual $\beta$ term be accompanied by an unphysical torque whose strength grows linearly with the amount of electric charge passing through the ferromagnetic wire. 

\subsection{Dissipative? Yes}
\label{sec:NASTT-is-dissipative}

The nonadiabatic spin-transfer torque is neither conservative, nor gyroscopic. The remaining possibility would be dissipative. 

The Rayleigh function for a one-dimensional ferromagnet respecting the symmetries of translation and global spin rotation is 
\begin{equation}
R = \frac{1}{2} r \dot{Q}^2 
+ \frac{1}{2} \alpha \mathcal S 
\int dx \, (\partial_t \mmm)^2
+ \beta \frac{\hbar}{e} \dot{Q}
\int dx \, 
\partial_t \mmm \cdot \partial_x \mmm
\end{equation}
to the lowest order in spatial gradients. The first term describes ohmic dissipation; $r$ is the resistance of the ferromagnetic wire. The second term gives rise to Gilbert damping in magnetization dynamics \cite{Gilbert:2004}; $\alpha>0$ is Gilbert's dimensionless damping constant. The third term, mixing the electric and ferromagnetic channels yields the nonadiabatic spin-transfer torque: 
\begin{equation}
\mathcal S \, \partial_t \mmm = 
- \mmm \times \frac{\delta U[\mmm]}{\delta \mmm}
- \frac{\hbar \dot{Q}}{2e} \partial_x \mmm
- \alpha \mathcal S \mmm \times \partial_t \mmm
- \beta \frac{\hbar \dot{Q}}{e} \mmm \times \partial_x\mmm.
\label{eq:LL-1d-torques-with-dissipation}
\end{equation}

The emf from dissipative forces is
\begin{equation}
\mathcal E_d
= - \frac{\partial R}{\partial \dot{Q}}
= - r \dot{Q} 
- \beta \frac{\hbar}{e}
\int dx \, 
\partial_t \mmm \cdot \partial_x \mmm.
\label{eq:emf-dissipative}
\end{equation}
The first term is the voltage due to ohmic losses, the second is the dissipative emf from spin dynamics. 

In the presence of dissipation, the unified equations of motion (\ref{eq:eom-collective-coordinates-with-Q}) read
\begin{equation}
- \frac{\partial W}{\partial q^\alpha}
+ G_{\alpha\beta} \dot{q}^\beta 
- \Gamma _{\alpha\beta} \dot{q}^\beta
= 0,
\end{equation}
where the purely magnetic dissipative coefficients are \cite{Clarke:2008}
\begin{equation}
\Gamma_{ij} = 
\alpha \mathcal S 
\int dx \, 
\frac{\partial \mmm}{\partial q^i}
\cdot
\frac{\partial \mmm}{\partial q^j}.
\end{equation}

For a domain wall (\ref{eq:domain-wall-solution}) in a ferromagnetic wire, the equations of motion for the four collective coordinates $X$, $\Psi$, $Q$, and $\Phi$ read
\begin{eqnarray}
X: &&
-\frac{\partial U}{\partial X}
\mp 2 \mathcal S \dot{\Psi} 
- 2\alpha \mathcal S \frac{\dot{X}}{\lambda}
+ 2\beta \frac{\hbar}{e} \frac{\dot{Q}}{\lambda} 
= 0,
\label{eq:eom-dw-X}
\\
\Psi: && 
-\frac{\partial U}{\partial \Psi}
\pm 2 \mathcal S \dot{X}
\mp \frac{\hbar}{e} \dot{Q}
- 2\alpha \mathcal S \lambda \dot{\Psi}
= 0,
\label{eq:eom-dw-Psi}
\\
Q: && 
V 
- \dot{\Phi}
\pm \frac{\hbar}{e}\dot{\Psi}
- r \dot{Q}
+ 2\beta \frac{\hbar}{e} \frac{\dot{X}}{\lambda} 
= 0,
\label{eq:eom-dw-Q}
\\
\Phi: &&
- \frac{\Phi}{\ell} + \dot{Q} = 0.
\label{eq:eom-dw-Phi}
\end{eqnarray}
The adiabatic spin-transfer torque manifests itself as the torque $\mp \frac{\hbar}{e}\dot{Q}$ on the domain wall in Eq.~(\ref{eq:eom-dw-Psi}) and the emf $\pm \frac{\hbar}{e}\dot{\Psi}$ in Eq.~(\ref{eq:eom-dw-Q}). The combined work of these two generalized forces vanishes, as it should for a gyroscopic force: 
\begin{equation}
\mp \frac{\hbar}{e}\dot{Q} \, d\Psi 
\pm \frac{\hbar}{e}\dot{\Psi} \, dQ 
= \mp \frac{\hbar}{e}
(\dot{Q}\dot{\Psi} - \dot{\Psi}\dot{Q})dt = 0.
\end{equation}
The nonadiabatic counterpart appears in  Eq.~(\ref{eq:eom-dw-X}) as the force $2\beta \frac{\hbar}{e} \frac{\dot{Q}}{\lambda}$ on the domain wall and in Eq.~(\ref{eq:eom-dw-Q}) as the emf $2\beta \frac{\hbar}{e} \frac{\dot{X}}{\lambda}$. Their combined work is nonzero: 
\begin{equation}
2\beta \frac{\hbar}{e} \frac{\dot{Q}}{\lambda} \, dX
+ 2\beta \frac{\hbar}{e} \frac{\dot{X}}{\lambda} \, dQ
= 2\beta \frac{\hbar}{e} 
\frac{\dot{Q} \dot{X} + \dot{X}\dot{Q}}{\lambda}dt \neq 0.
\end{equation}

\section{Discussion}
\label{sec:discussion}

Expressing the dynamics of ferromagnetic solitons in terms of collective coordinates \cite{Thiele:1973, Tretiakov:2008} divides the forces acting on a soliton into three categories: conservative, dissipative, and gyroscopic. The first type is independent of soliton velocities, the other two are linear in them, with symmetric or antisymmetric proportionality coefficients. Viewed from this limited perspective, the spin-transfer torques do not depend on velocities of a soliton and at first glance appear to be conservative. However, this impression is misleading. We have given two clear-cut examples where the work of spin-transfer torques over a closed loop is nonzero, so this force is not conservative. 

Expanding the physical system to include the electric charge $Q$ as a dynamical variable leads to a realization that the spin-transfer torques are proportional to the charge velocity $\dot{Q}$, which is characteristic of gyroscopic and dissipative forces. We have shown that the adiabatic spin-transfer torque is a gyroscopic force and obtained the geometric action from which it can be derived. 

The nonadiabatic spin-transfer torque cannot be described as a gyroscopic force. An attempt to do so fails because of strict limitations on the structure of the gyroscopic force: a gyroscopic spin-transfer torque includes an unphysical term whose magnitude is proportional to the amount of charge $Q$ that has flown through the ferromagnet. The nonadiabatic spin-transfer torque is thus not a gyroscopic but dissipative force. This conclusion is consistent with the argument from Onsager's reciprocity principle \cite{Brataas:2017}, which requires that the coefficients $\beta$ in the nonadiabatic spin-transfer torque (\ref{eq:LL-1d-NASTT-torques}) and in the spin-induced emf (\ref{eq:emf-dissipative}) be the same. A gyroscopic component would add contributions of equal size but opposite sign to the two $\beta$ coefficients. 

Our conclusions are based on the trichotomy of forces---conservative, dissipative, and gyroscopic---well-established for ferromagnetic solitons and extended here to electric charge. It is possible that this classification misses some category of forces. Indeed, it omits the forces of inertia linear in accelerations and associated with kinetic energy. Inertial forces are not \emph{fundamental} for ferromagnetic solitons: they do not exist at the starting point, the Landau--Lifshitz equation (\ref{eq:LL-1d-torques-with-dissipation}). However, inertia can \emph{emerge} from a combination of conservative and gyroscopic forces as a result of a coarse-graining procedure (the D{\"o}ring mass \cite{Doring:1948, Rado:1950, Rado:1951}). At any rate, the spin-transfer torques are clearly not inertial. Barring the identification of yet another force category relevant to magnets (which would be an interesting story in its own right), we hope that our paper clears up at least some of the lingering questions about the nature of the spin-transfer torques. 

For the sake of simplicity, we limited the discussion to ferromagnets in one spatial dimension. Extension to higher dimensions brings up some interesting questions. E.g., how does one define charge variables in higher dimensions? Charge conservation turns out to play an essential role, so situations in which the electric current enters and leaves a ferromagnet through nonmagnetic leads require careful handling. 

To illustrate that, we note that the deceptively simple example of a domain wall in a ferromagnetic wire used in Sec.~\ref{sec:NASTT-is-dissipative} contains an inherent contradiction. In deriving its equations of motion (\ref{eq:eom-dw-X}--\ref{eq:eom-dw-Phi}), we assumed that the ferromagnetic conductor forms a loop; otherwise, the magnetic flux variable $\Phi$ would be ill-defined. However, it is not possible to have a single domain wall in a ferromagnetic loop, for a loop must contain an even number of domain walls. A simple way out would be to consider two well-separated and non-interacting domain walls in a ferromagnetic loop. Eqs.~(\ref{eq:eom-dw-X}--\ref{eq:eom-dw-Phi}) can be readily upgraded to that case. 

A more interesting and challenging problem would be to consider a loop made of two connected wires, ferromagnetic and nonmagnetic. We can then consider a single domain wall in the ferromagnetic part. The challenge would be to describe the gradual loss of spin polarization happening when the electric current leaves the ferromagnetic wire and enters the nonmagnetic one. Staying away from microscopic models of spin relaxation in order to keep things simple, one could introduce thermal noise inducing Brownian motion of all physical variables along the lines of Tserkovnyak and Wong \cite{Tserkovnyak:2009}. In the nonmagnetic wire, the magnetization field would have zero energy and would therefore be quickly randomized by thermal fluctuations. 

We hope to address these issues in a subsequent publication. 

\section*{Acknowledgement}

We thank Sayak Dasgupta, Paul Haney, Allan MacDonald, and Mark Stiles for very helfpul discussions. This work was supported as part of the Institute for Quantum Matter, an Energy Frontier Research Center funded by the U.S. Department of Energy, Office of Science, Basic Energy Sciences under Award No. DE-SC0019331.

\appendix

\section{Circuit theory from the Aharonov--Bohm phase}
\label{app:electric-circuit-geometric-phase}

To support the heuristic arguments advanced in the main text, we provide a formal derivation of the spin-transfer torque in a very simple model of a conducting ferromagnet. This derivation leaves out the microscopic details such as electron interactions and focuses directly on the underlying cause of the gyroscopic force, namely the geometric action, or the spin Berry phase. 

\begin{figure}
    \centering
    \includegraphics[width=0.8\columnwidth]{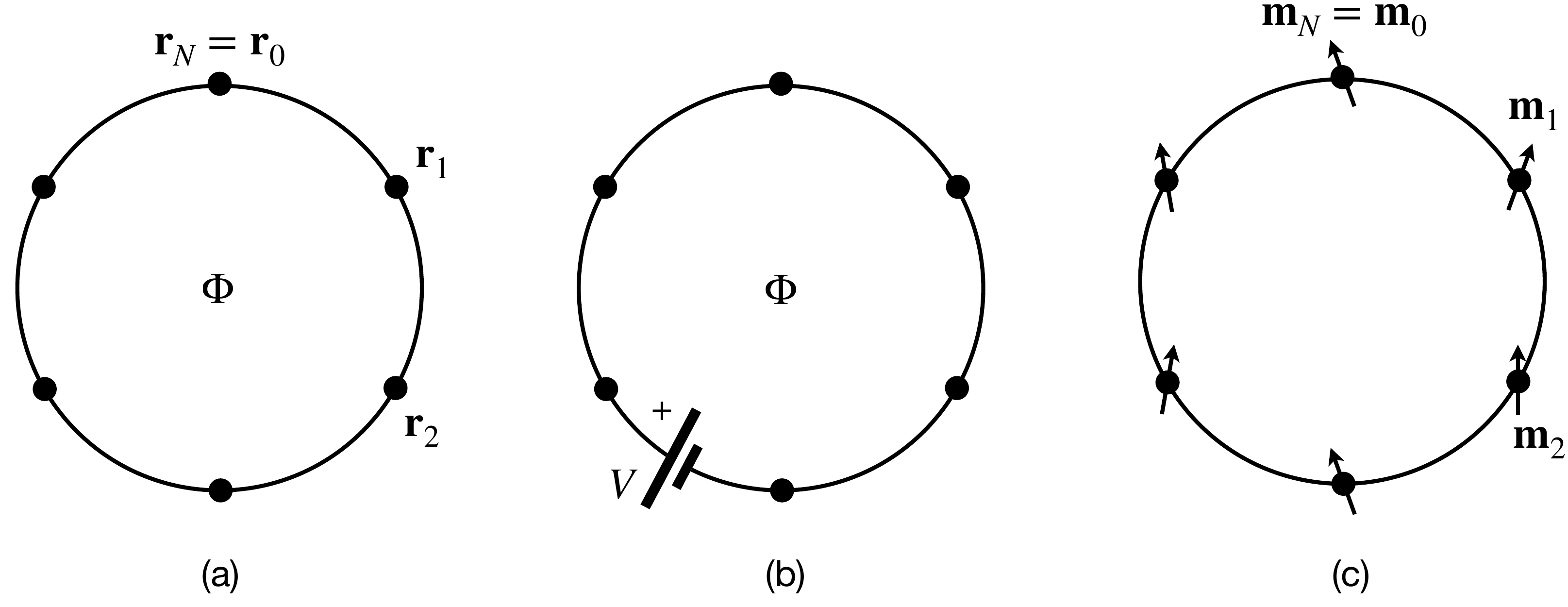}
    \caption{(a) An electric circuit made of small metallic grains contains a magnetic flux $\Phi$. (b) The circuit is endowed with a battery (emf $V$). (c) The grains are made magnetic.}
    \label{fig:circuits}
\end{figure}

We start by deriving the equations of motion for an electric circuit. We represent the circuit as a discrete set of small metallic grains located at positions $\rrr_1, \rrr_2, \ldots, \rrr_N = \rrr_0$, Fig.~\ref{fig:circuits}(a). A high charging energy enforces electric neutrality for each grain so that, when an electron with charge $e$ jumps from grain $1$ to grain $2$, another electron has to leave grain $2$ for grain $3$, etc. Thus an elementary move consists of transferring one electron from each grain $n$ to its neighbor $n+1$. 

If the circuit contains a magnetic flux $\Phi$, the electrons will collectively acquire the Aharonov-Bohm phase 
\begin{equation}
\phi^\mathrm{AB} 
= \frac{e}{\hbar} \sum_{n=1}^N \AAA(\rrr_n) \cdot \Delta \rrr_n 
\approx \frac{e}{\hbar} \oint \AAA(\rrr) \cdot d \rrr
= \frac{e \Phi}{\hbar}.
\end{equation}
where $\Delta \rrr_n = \rrr_{n+1} - \rrr_n$ and $\AAA(\rrr_n)$ is the gauge potential at the location of grain $n$. 

We may now interpret the elementary charge $e$ as an increment of the physical variable $Q(t)$ counting the amount of charge that passes through any given grain since the time $t=0$. The Aharonov-Bohm phase translates to a geometric action 
\begin{equation}
S_g^\mathrm{AB} = \hbar \phi^\mathrm{AB} = \int \Phi \, dQ
= \int \Phi \dot{Q} \, dt.
\end{equation}
Hence the geometric part of the Lagrangian for the electric circuit $L_g^\mathrm{AB} = \Phi \dot{Q}$. 

To make our electric circuit a bit more realistic, we add to it a battery with emf $V$, Fig.~\ref{fig:circuits}(b), and endow the magnetic flux with magnetostatic energy $U^\mathrm{MS} = \Phi^2/2\ell$, where $\ell$ is the self-inductance of the circuit. Its Lagrangian is then 
\begin{equation}
L = \Phi \dot{Q} + V Q - \frac{\Phi^2}{2\ell}.    
\end{equation}
It is easy to check that this Lagrangian yields the equations of motion for the electric charge (\ref{eq:eom-Q}) and magnetic flux (\ref{eq:eom-Phi}), with the exception of the gyroscopic coupling of the charge to the ferromagnet. 

\begin{figure}
    \centering
    \includegraphics[width=0.9\columnwidth]{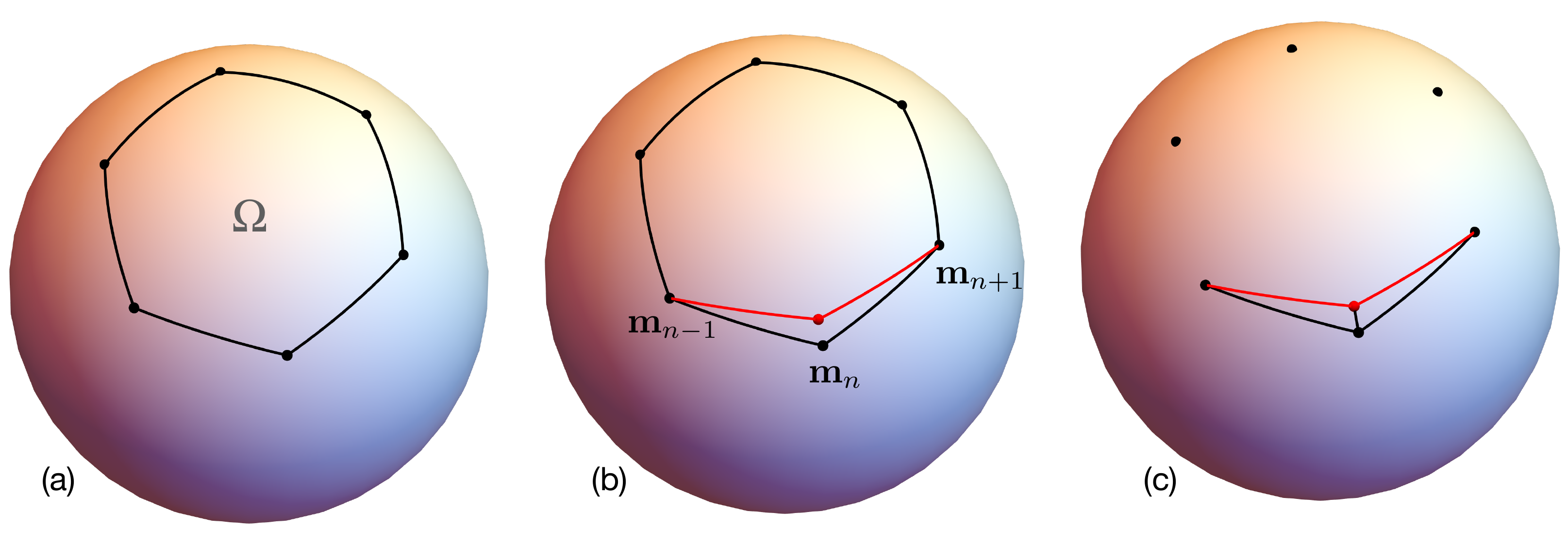}
    \caption{(a) A path on the unit sphere. }
    \label{fig:paths-on-sphere}
\end{figure}

\section{Adiabatic spin-transfer torque from the spin Berry phase}
\label{app:ferromagnet-geometric-phase}

We now make the grains ferromagnetic with perfect spin polarization: all electrons in grain $n$ have spin pointing in a common direction $\mmm_n$, Fig.~\ref{fig:circuits}(c). An electron moving from grain $n$ to grain $n+1$ then changes its spin from $\hbar \mmm_n/2$ to $\hbar \mmm_{n+1}/2$. Collectively, the electron spins trace a closed path on the unit sphere of magnetization connecting points $\mmm_1, \mmm_2, \ldots, \mmm_N = \mmm$, Fig.~\ref{fig:paths-on-sphere}(a). The net Berry phase acquired by the electron spins is $-\Omega/2$, where $\Omega$ is the oriented area delimited by the path. The geometric action for one elementary charge move, $\delta Q = e$, is $S_g^\mathrm{ASTT} = - \hbar \Omega/2$. Integrating the action over many elementary moves yields
\begin{equation}
S_g^\mathrm{ASTT} = - \frac{\hbar}{2e} \int \Omega \, dQ    
= - \frac{\hbar}{2e} \int \Omega \dot{Q} \, dt.
\end{equation}
The emf arising from this term is 
\begin{equation}
\mathcal E^\mathrm{ASTT} = \frac{\delta S_g^\mathrm{ASTT}}{\delta Q}
= \frac{\hbar \dot{\Omega}}{2e},
\end{equation}
which agrees, as expected, with Eq.~(9) of Barnes and Maekawa \cite{Barnes:2007}. 

To obtain the spin-transfer torque, we need to compute the functional derivative 
\begin{equation}
\frac{\delta S_g^\mathrm{ASTT}}{\delta \mmm_n} 
= - \frac{\hbar \dot{Q}}{2e} \frac{\partial \Omega}{\partial \mmm_n}.
\end{equation}
The change of the area $\delta \Omega$ in response to an infinitesimal shift $\delta \mmm_n$, Fig.~\ref{fig:paths-on-sphere}(b), is equal to the combined area of two spherical triangles containing vertices $\mmm_n$ and $\mmm_n + \delta \mmm_n$, Fig.~\ref{fig:paths-on-sphere}(c). With the aid of the formula for the area of a spherical triangle with vertices $\mmm_1$, $\mmm_2$, and $\mmm_3$ \cite{Eriksson:1990},
\begin{equation}
\tan{\frac{\Omega}{2}}
= \frac{\mmm_1 \cdot (\mmm_2 \times \mmm_3)}
{1 + \mmm_2 \cdot \mmm_3 + \mmm_3 \cdot \mmm_1 + \mmm_1 \cdot \mmm_2},
\end{equation}
we obtain the derivative
\begin{equation}
\frac{\partial \Omega}{\partial \mmm_n}
= \frac{\mmm_{n-1} \times \mmm_n}
{1 + \mmm_{n-1} \cdot \mmm_n}
+ \frac{\mmm_{n+1} \times \mmm_n}
{1 + \mmm_{n+1} \cdot \mmm_n}.
\end{equation}
Hence the spin-transfer torque for magnetic grain $n$,
\begin{equation}
\mmm_n \times \frac{\delta S_g^\mathrm{ASTT}}{\delta \mmm_n}   
= \frac{\hbar \dot{Q}}{2e} \mmm_n \times
\left(
\frac{\mmm_{n-1} \times \mmm_n}
{1 + \mmm_{n-1} \cdot \mmm_n}
- \frac{\mmm_n \times \mmm_{n+1} }
{1 + \mmm_n  \cdot \mmm_{n+1}}
\right).
\label{eq:STT-m-n}
\end{equation}
This expression exactly reproduces the standard form of discrete the spin-transfer torque for a fully polarized ferromagnet, Eq.~(5) of Slonczewski \cite{Slonczewski:1999}. 

We may pass to the continuum model of a ferromagnet (in 1 dimension) by setting $\mmm_n = \mmm(x)$ and using the gradient expansion, $\mmm_{n\pm1} = \mmm(x\pm a) \approx \mmm(x) \pm a \partial_x \mmm(x)$, where $a$ is the lattice constant. The discrete spin-transfer torque (\ref{eq:STT-m-n}) then reduces to its continuum counterpart (\ref{eq:STT-m}). 

\bigskip

\bibliographystyle{iopart-num}
\bibliography{main}

\end{document}